\documentclass[12pt,showpacs,preprintnumbers]{revtex4}

\usepackage{graphicx}
\usepackage{dcolumn}
\usepackage{bm}
\usepackage{amssymb}
\usepackage{epsfig}    
\usepackage{here}
\def\beq{\begin{equation}}
\def\eeq{\end{equation}}
\def\ba{\begin{array}}
\def\ea{\end{array}}
\def\bea{\begin{eqnarray}}
\def\eea{\end{eqnarray}}

\def\sq2{\sqrt{2}}
\def\End{\end{document}}

\def\Journal#1#2#3#4{{#1} {\bf #2} (#4) #3}

\def\PLB{{\em Phys. Lett.}  B}

\def\PRD{{\em Phys. Rev.} D}

\begin{document}                                                              

\title{General Analysis of Single Top Production and \\
W Helicity in Top Decay}%
\author{%
{\sc Chuan-Ren Chen\,$^1$,~~  F. Larios\,$^{1,2}$},~~    
{\sc C.--P. Yuan\,$^1$}
}
\affiliation{%
\vspace*{2mm} 
$^1$Department of Physics and Astronomy, 
Michigan State University, East Lansing, Michigan 48824, USA\\
$^2$Departamento de F\'{\i}sica Aplicada,
CINVESTAV-M\'erida, A.P. 73, 97310 M\'erida, Yucat\'an, M\'exico
}

\begin{abstract}
\hspace*{-0.35cm}
We provide a framework for the analysis of the $W$ boson helicity
in the decay of the top quark that is based on a general effective
$tbW$ coupling.  Four independent coupling coefficients can be
uniquely determined by the fractions of longitudinal and transverse
$W$ boson polarizations as well as the single top production rates
for the t-channel and the s-channel processes.
The knowledge of these coefficients can be used to discriminate
models of electroweak symmetry breaking.
\pacs{\,12.60.-i,\,12.15.-y,\,11.15.Ex,\,14.65.Ha,\,14.70.Fm 
\hfill   ~~ [ March, 2005 ] }

\end{abstract}

\maketitle

\setcounter{footnote}{0}
\renewcommand{\thefootnote}{\arabic{footnote}}

\section{Introduction}
The top quark stands out as the heaviest elementary particle known
to date.  It lives very shortly and almost all of the time decays
into a $b$ quark and a $W$ boson~\cite{review}.
Because of the top quark mass being of order the
electroweak symmetry breaking (EWSB) energy scale, studying the top quark
interactions is of great interest.  The knowledge of these interactions
is required in order to discriminate mechanisms of
EWSB. Moreover, because of the top's decay mode $t\to bW$,
the $tbW$ coupling plays a significant role in the physics of the top quark.

One of the main goals at the Fermilab Tevatron and at the CERN Large
Hadron Collider (LHC) is to study the production and decay of top quarks.
The measurement of single top production cross section has
turned out to be a challenging task and no single top events have been
observed so far~\cite{reinhard}.
This non-observation is translated into upper limits of order 5 pb 
(based on $230\,{\rm pb}^{-1}$ integrated luminosity)
for
each production channel~\cite{reinhard}, far above the predictions of the
Standard Model (SM) which are of order $1-2$ pb.  However, it is expected
that more luminosity and improved analysis methods will eventually
achieve detection of SM single top events.

There are
three modes in the $t\to bW$ decay, 
depending on the polarization state of the $W$ boson.
Each mode is associated with a fraction,  $f_0$, $f_+$ or $f_-$,
that corresponds to the longitudinal, right-handed or left-handed
polarization, respectively.
By definition, we have the constraint $f_0+f_++f_-=1$.
Recent reports by the D\O{} and CDF collaborations at Fermilab give
the following ($95 \%$ C.L.) results for the longitudinal and
right-handed fraction of $t\to bW$ in the $t\bar t$ pair
events~\cite{exptpol}:
\bea
f_0 &=& 0.91 \pm 0.38\; ({\rm CDF})\, ,\;\;\;\;
f_0\; = \;0.56\;\pm 0.32 \; ({\rm D\O{}})\, ,
\nonumber \\
f_+ &\le& 0.18 \; ({\rm CDF})\, ,\hspace{1.7cm}
f_+\; \le \;0.24\; ({\rm D\O{}})\, .
\nonumber
\eea
In this work we propose a new strategy to use the measurements
on the single top production cross section and on the polarization
of the $W$ boson in the $t\to bW$ decay in order to determine the
general effective $tbW$ vertex.   Our strategy consists of using four
measurements: a) $\sigma_s$ and $\sigma_t$, the cross sections of
the two most important modes of single top quark production at the
Tevatron, referred to as s-channel and t-channel~\cite{sullivan},
 and 
b) two of the three decay ratios, $f_0$, $f_-$ and $f_+$, to determine
the four independent couplings that define the general effective
$tbW$ vertex.  To emphasize the importance of measuring the $tbW$
vertex, we will consider two different models of EWSB, and compare
their predictions on $tbW$.  In this manner, we show that the proposed
analysis can help us to distinguish different models of EWSB.

\section{The general approach to study top quark interactions}
Currently, the only missing ingredient of the SM is the Higgs boson.
This is the agent that causes the breaking of the electroweak symmetry,
and LEPII searches have concluded that its mass must be greater than
$115\,$GeV if such particle exists~\cite{leph}.  It is well known that the
Higgs mechanism in the SM leaves many important questions unanswered;
like what is the real origin of the fermion masses, or what is the
explanation for a significant cancellation of higher order corrections
to the Higgs mass.  As a result, other theories of EWSB are given much
attention in the particle physics community.  Theories like the
Minimal Supersymmetric Standard Model (MSSM), the Technicolor models,
and theories with new top quark interactions suggest some of the
answers, but so far no indication of their validity has been found.

Another approach to study the physics that is responsible for EWSB is
to focus our attention on the particles that we know exist.
Whatever new physics interactions
may exist, they must become apparent at an energy scale higher
than what we have been able to probe so far.  We do not know how
high this scale may be.  Maybe it lies much higher than the electroweak
scale ($246\,$GeV) and if so, the only way we can begin to get information
about these interactions is by looking at the effects they produce on
the interactions appearing at lower energies.   Because of their big
masses, the top quark, the $W$ and the $Z$ bosons are the prime
candidates to show these effects through their interactions.

In this paper we want to provide a general framework that describes
all the possible effects from any physics beyond the SM.
This framework is based on the non-linear electroweak chiral 
Lagrangian~\cite{nonlinear}.  This Lagrangian satisfies the
$SU(2)_L\times U(1)_Y$ symmetry by a non-linear realization, and
it is the most general Lagrangian that is consistent with the SM
gauge symmetry and that can contain all the possible effects
(decoupled and non-decoupled) coming from the physics at higher
energy scales.  Concerning the $tbW$ system, it has been shown that
the leading dimension 4 and dimension 5 interaction terms that are
independent from each other are~\cite{operators}:
\bea
{\cal L}_{(tbW)} = -\frac{1}{\sqrt{2}} \left( (1+\kappa_{L}^{(4)})
\bar t \gamma^{\mu} P_L { b} + \kappa_{R}^{(4)}
\bar t \gamma^{\mu} P_R {b} \right) {{\cal W}_{\mu}^+}
\hspace{1cm}
\nonumber \\
+ {\frac {\kappa_{R(L)}^{\sigma}}{\Lambda}} \; \bar b
\sigma^{\mu \nu} P_{R(L)} t \, D_{\mu} {\cal W}^{-}_{\nu} 
+ i{\frac {\kappa_{R(L)}^{t}}{\Lambda}} \; {\cal W}^{-\mu}
\bar b P_{L(R)} \, D_{\mu} t  \hspace{1.0cm}
\nonumber  \\
+ i{\frac {\kappa_{R(L)}^{w}}{\Lambda}} \;
\bar b P_{R(L)} t \, D^{\mu} {\cal W}^{-}_{\mu} \;\;\;\;\;
+ \; h.c.\, , \hspace{1cm}
\nonumber
\eea
where $P_{R(L)}$ are the right-and left-handed chiral projectors
$P_{R(L)}=(1\pm\gamma_5)/2$, $D_\mu$ is the electromagnetic $U(1)$
covariant derivative and $\Lambda$ is the energy scale at
which the physics beyond the SM becomes apparent.
The $t$, $b$ and $\cal W^+$ fields are not the usual fermion and
vector boson fields.  Rather, they are {\it composite} fields that
involve Goldstone boson fields and that transform non-linearly under
the gauge group~\cite{operators}.  In the unitary gauge they become
the usual fields (e.g., ${\cal W}^+ = -g W^+$).
In the remainder of this letter, $t$ and $b$ denote the usual 
fermion fields for the top and bottom quarks.  
To simplify our analysis, the $\kappa$ coefficients are taken to
be real so that there are no CP violation effects.

The effective $tbW$ coupling generated by this Lagrangian contains
terms proportional to $\gamma_\mu$, $\sigma_{\mu\nu} q^\nu$, $p_\mu$
and $q_\mu$, with $p$ and $q$ the momenta of the top quark and the
$W$ boson, respectively.   We can make a simplification of this
vertex that is valid for our study.
First of all, since the $t\to bW$ decay involves quarks on-shell,
we can use the well known Gordon identity:
\bea
(m_b+m_t) \bar b \gamma_\mu t = \bar b
(p_\mu+{p'}_\mu-i\sigma_{\mu\nu}q^\nu) t \, ,
\nonumber
\eea
where ${p'}=p-q$ is the momentum of the $b$ quark, and reduce
the degrees of freedom to three terms:  
$\gamma_\mu$, $\sigma_{\mu\nu} q^\nu$ and $q_\mu$.  Because
of the on-shell condition of the $W$ boson, the term proportional
to $q_\mu$ will not contribute to the $t\to bW$ decay amplitude.
Furthermore, this $q_\mu$ term will neither contribute to the
single top production processes, because it will only generate a
contribution proportional to the incoming state light quark
masses which are usually taken as zero.

Therefore, the effects of our general effective Lagrangian to the
processes considered here can be completely described by the
following $tbW$ vertex:
\bea
\mathcal{L}_{\mathrm tbW} &=&  \frac{g}{\sqrt 2}\, W^-_\mu \, 
\bar b \, \gamma^\mu  \left( f_1^L P_L + f_1^R P_R  \right)\, t
\nonumber \\
&-& \frac{g}{\sqrt 2 M_W} \,
\partial_\nu W^-_\mu \, \bar b \, \sigma^{\mu\nu}
\left( f_2^L P_L + f_2^R P_R \right) \, t \;\;\;+\; h.c.\, ,
\label{tbwvertex} 
\eea
where we have changed the mass scale $\Lambda$ to $m_W$ to keep
the same notation used in the literature~\cite{kane,aguila}.

In the SM the values of the form factors are
$f^L_1 = V_{tb}\simeq 1$, $f^R_1 = f^L_2 = f^R_2 = 0$.
To focus on deviations from SM values, let us define
$f_1^L \, \equiv \, 1\, +\,\epsilon_L$.

It is well known that $b\to s \gamma$ can impose a strong constraint
on $f_1^R$ and $f_2^L$ to be less than $0.004$~\cite{perez,burdman}.
These constraints can be viewed as the result of an $m_b$
suppression for right-handed bottom quark couplings~\cite{burdman}.
On the other hand,  $b\to sl^+l^-$ can be sensitive to a 
left-handed bottom quark coupling like $f_2^R$, and it can impose a
constraint of order $0.03$~\cite{burdman}.
For $\epsilon_L$, the LEP precision data imposes some constraint
but only in correlation with similar neutral current anomalous
$ttZ$ couplings.  Assuming no deviations from the SM $ttZ$ vertex
we would have that $\epsilon_L \leq 0.02$~\cite{perez}.
To bear in mind, these constraints assume there are no other sources
of new physics that could cancel the effects of these couplings on
the data.  Moreover, the dimension 5 couplings $f_2^R$
and $f_2^L$ may induce a bad high energy behavior in top quark
production processes, hence, we will consider values at most of
order $0.5$ in order to satisfy the unitarity condition~\cite{renard}. 


Studies of the dimension 5 couplings $f_2^{L,R}$  in connection with
the single top quark production at hadron colliders have shown that
a sensitivity of order 0.2 (0.05) might be achieved at the Tevatron
(LHC)~\cite{boos}.
Information on the helicity of the $W$ boson in $t\to bW$ can
be obtained by measuring a forward-backward asymmetry ($A_{FB}$)
based on the angle between the charged
lepton and the b-jet of the observed decay process~\cite{korner}.
Preliminary studies show that if $A_{FB}$ is measured with $20\%$
accuracy at the Tevatron, it may be sensitive to values of order
$f_2^{L,R} \sim 0.3$;  similarly, if $A_{FB}$ is measured with
$1\%$ accuracy at the LHC this may be translated to a sensitivity
of order $f_2^L\sim 0.03$ and $f_2^R\sim 0.003$~\cite{aguila}.

We would like to point out that, since the observable $A_{FB}$ is
only proportional to the difference between $f_+$ and
$f_-$~\cite{korner}, it is clear that it does not provide any
more information than the separate measurements of (two of) the
ratios $f_0$, $f_-$ and $f_+$.

Let us summarize the status of the SM predictions for the
observables of our study: the cross sections $\sigma_t$ and
$\sigma_s$, and the branching fractions $f_0$, $f_+$ and $f_-$.
In Table~\ref{smsingletop} we show the leading order (LO) and
the next-to-leading order (NLO) SM predictions for $\sigma_t$ and
$\sigma_s$ at the Tevatron and at the LHC~\cite{sullivan}.
For the LO predictions the CTEQ6L1 parton distribution function
(PDF) has been used~\cite{pdf}.  For the NLO predictions the
CTEQ6M PDF has been used~\cite{sullivan}.  In this letter we
are taking the mass of the top quark as $m_t=178\,$GeV
and the mass of the $W$ boson as $m_W=80.4\,$GeV.

\begin{table}[ht]
\begin{tabular}{|c||c|c||c|c||c|c|}
\hline Channel & Tevatron ($t$ {\small LO}) & ($t$ {\small NLO}) &
LHC ($t$ {\small LO}) & ($t$ {\small NLO}) &
LHC ($\bar t$ {\small LO}) & ($\bar t$ {\small NLO}) 
\tabularnewline \hline \hline
t-channel & 0.827 & 0.924 & 146.0 & 150.0 & 84.9 & 88.5 
\tabularnewline \hline 
s-channel & 0.27 & 0.405 & 4.26 & 6.06 & 2.59 & 3.76
\tabularnewline \hline
\end{tabular}
\caption{SM single top production cross section predictions in units
of pb~\cite{sullivan}.
The mass of the top quark is taken as $m_t=178\,$GeV.
\label{smsingletop}}
\end{table}

Neglecting terms proportional to the bottom mass, the Born level
values of the top quark width and its $W$-polarization ratios are
$\Gamma_t=1.65\,$GeV, $f_0=0.71$, $f_-=0.29$ and $f_+ =0$.
In the SM, including terms proportional to $m_b$, order $\alpha_s^2$ QCD,
electroweak, and finite $W$ width corrections produce a $10\%$
decrease in the top's width ($\Gamma_t = 1.49$) and a small $\sim 1\%$
variation for decay ratios
($f_0 = 0.701$, $f_- = 0.297$ and $f_+ = 0.002$)~\cite{korner}.

In this work we will be interested in deviations from the SM values
(up to the NLO) that come from the effects of the anomalous
$\epsilon_L$, $f_1^R$, $f_2^L$ and $f_2^R$ couplings, 
cf. Eq.~(\ref{tbwvertex}), induced by heavy new physics effects.
In the following, we will write down the Born level contributions
of these couplings on the observables $f_0$, $f_+$, $f_-$, $\sigma_t$
and $\sigma_s$.

\section{Single top production and $W$ helicity in $t\to bW$ decay}
The tree level $t\to b W$ decay width of the top quark with the
general $tbW$ vertex can be easily obtained with the helicity
amplitude method, and it is given by~\cite{kane}:
\begin{eqnarray}
\Gamma_t &=& \Gamma_0 + \Gamma_- +\Gamma_+ \nonumber \\
&=& \frac{g^2 m_t}{64\pi} \; \frac{(a_t^2-1)^2}{a_t^4}
\left( a_t^2 L_0^2 + 2T_m^2 + 2T_p^2 \right)\, ,
\nonumber \\
L_0^2 &\equiv& 1+x_0\; =
(f_1^L + f_2^R/a_t)^2 + (f_1^R + f_2^L/a_t)^2 \, ,
\nonumber \\
T_m^2 &\equiv& 1+x_m\; =(f_1^L + a_t f_2^R)^2 \, ,
\nonumber \\
T_p^2 &\equiv& x_p\; = (f_1^R + a_t f_2^L)^2\, ,
\label{width} \\
a_t &\equiv& \frac{m_t}{m_W}\, . \nonumber
\end{eqnarray}
As the notation suggests, $x_0$, $x_m$ and $x_p$ are the effective
terms that originate the contribution to $f_0$, $f_-$ and $f_+$,
respectively.  Below, we will write down the explicit
expressions for these decay ratios.

The t-channel total cross section at the parton level comes from
two processes: $ub\to dt$ and $\bar d b\to \bar u t$.  For the
first one the expression is:
\begin{eqnarray}
\sigma (ub\to dt) &=& \frac{g^4}{64 \pi s}
( I_0 L^2_0 +  I_m T^2_m +  I_p T^2_p - I_i x_i +  I_5 x_5 )
\, , \nonumber \\
I_0 &=& x_t (C_b - C_a) \, , \nonumber \\
I_m &=& C_a - x_t C_b \, , \label{sectiont} \\
I_p &=& I_m + (1+C_{tw}) (x_w C_a -C_l) +1- x_t -x_w C_l
\, , \nonumber \\
I_i &=& (\ln x_t + C_{tw} C_l )/( x_t - x_w) \, , \nonumber \\
I_5 &=& 1 - (1+\ln x_t)/x_t - 2I_i/a^2_t \, , \nonumber \\
x_5 &=& a_t^2 ({f_2^L}^2+{f_2^R}^2) \, , \nonumber \\
x_i &=& 2a_t ( {f_1^L}{f_2^R} + {f_2^L}{f_1^R} ) \,=\,
\frac{a^2_t}{a^2_t-1} (x_m+x_p-x_0) -
\frac{1+a^2_t}{a^2_t} x_5 \, , \nonumber  
\end{eqnarray}
where $s=(p_u+p_b)^2$ is the total energy squared of the colliding
partons.  We have defined the following terms:
\begin{eqnarray}
x_t = \frac{m^2_t}{s} \;\; ,\;\;\;\; x_w = \frac{m^2_w}{s}
\;\; ,\;\;\;\; C_{tw} = 1-x_t+x_w \;\; ,\;\;\;\;
C_l = \ln \frac{C_{tw}}{x_w} \;\; ,
\nonumber \\
C_a = \frac{1-x_t}{x_w C_{tw}} \;\; ,\;\;\;\;\;\; 
C_b = \frac{C_a}{x_t-x_w}
- \frac{C_l + \ln x_t}{(x_t - x_w)^2} \;\; .\;\;\;
\nonumber
\end{eqnarray}
The formula for $\bar d b\to \bar u t$ can be obtained from
Eq.~(\ref{sectiont}) by interchanging the coupling coefficients
$f_1^L \leftrightarrow f_1^R$ and $f_2^L \leftrightarrow f_2^R$
(or simply, $T_m^2 \leftrightarrow T_p^2$).  For the anti-top
production we have
$\sigma(\bar u \bar b\to \bar d \bar t)\,=\,\sigma(\bar d b\to \bar u t)$
and $\sigma(d \bar b\to u \bar t)\,=\,\sigma(u b\to d t)$.

The s-channel total cross section at the parton level is:
\begin{eqnarray}
\sigma (u\bar d\to t\bar b) &=& \frac{g^4}{128 \pi s} \,
 \frac{(s-m_t^2)^2}{(s-m_t^2)^2+m_w^2 \Gamma_w^2} \,
( T^2_m +  T^2_p - I_s ) \, ,\nonumber \\
I_s &=& ({f_1^L}^2+{f_1^R}^2 -x_5/x_t) (1-x_t)/3
\, .\label{sections} 
\end{eqnarray}
Where $\Gamma_w\,=2.1$ GeV is the $W$~boson's width.
The cross section formula for $u\bar d\to t\bar b$ is the same as
above.  To write Eq.~(\ref{sections}) in terms of the variables
$x_0$, $x_m$, $x_p$ and $x_5$, we can use the relation:
${f_1^L}^2+{f_1^R}^2 \,=\, 1+x_m+x_p-x_5-x_i$.

In summary, the contributions of the effective $tbW$
couplings to the observables of interest are:
\begin{eqnarray}
f_0 &=& \frac{a^2_t (1+x_0)}{a^2_t(1+x_0)+2(1+x_m+x_p)}\, ,
\label{eqf0} \\
f_+ &=& \frac{2x_p}{a^2_t(1+x_0)+2(1+x_m+x_p)}\, ,
\label{eqfp} \\
f_- &=& \frac{2(1+x_m)}{a^2_t(1+x_0)+2(1+x_m+x_p)}\, ,
\nonumber \\
\Delta \sigma_t &=& a_0 x_0 +a_m x_m + a_p x_p + a_5 x_5\, ,
\label{sigmat} \\
\Delta \sigma_s &=& b_0 x_0 +b_m x_m + b_p x_p + b_5 x_5\, ,
\label{sigmas} 
\end{eqnarray}
where $\Delta \sigma$ stands for the variation from the SM NLO
prediction.  The numerical values of the $a_i$ and $b_i$ coefficients
are given in Table~\ref{coefs} for the Tevatron and the LHC.
They have been obtained by integrating over the parton luminosities
which are evaluated using the PDF CTEQ6L1~\cite{pdf}.

\begin{table}[ht]
\begin{tabular}{|c||c|c|c|c|}
\hline t-channel: & $a_0$ & $a_m$ & $a_p$ & $a_5$
\tabularnewline \hline \hline
Tevatron & 0.896 & -0.069 & -0.153 & 0.292
\tabularnewline \hline 
LHC ($t$) & 165.2 & -19.1 & -34.2 & 71.7
\tabularnewline \hline 
LHC ($\bar t$) & 105.8 & -20.9 & -12.5 & 44.5 
\tabularnewline \hline \hline
s-channel: & $b_0$ & $b_m$ & $b_p$ & $b_5$
\tabularnewline \hline \hline
Tevatron & -0.081 & 0.352 & 0.352 & 0.230
\tabularnewline \hline 
LHC ($t$) & -1.41 & 5.67 & 5.67 & 6.34
\tabularnewline \hline 
LHC ($\bar t$) & -0.836 & 3.43 & 3.43 & 3.38
\tabularnewline \hline
\end{tabular}
\caption{The single top production cross section coefficients
of Eqs.~(\ref{sigmat}-\ref{sigmas}).  In units of pb.
\label{coefs}}
\end{table}

Eqs.~(\ref{eqf0})-(\ref{sigmas}) can be used to make a general
analysis of the effective $tbW$ vertex.  We note that in case a
new light resonance is found, like a scalar or vector boson,
the s-channel process could be significantly enhanced and its
production rate may not be dominated by a virtual $W$-boson
s-channel diagram~\cite{diaz}.

\begin{figure}
\includegraphics[width=9.2cm,height=7.4cm]{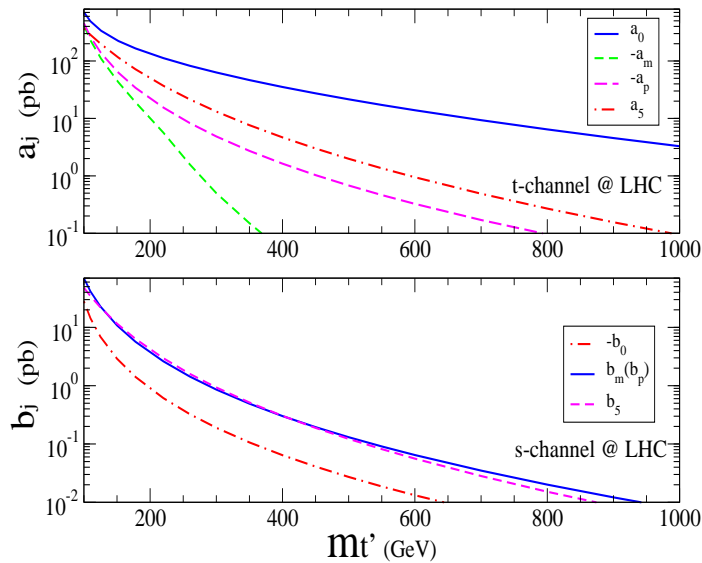}
\caption{The coefficients for the s and t channels of single
$t'$ production as given by
Eqs.~(\ref{sigmat}) and (\ref{sigmas}) at the LHC.}
\label{coefficient}
\end{figure}

The above formulas (summarized in Eqs.~(\ref{eqf0}-\ref{sigmas}))
also apply to models with extra heavy fermion ($t'$), such as the Little Higgs
Models \cite{perelstein}, that couples to the SM $b$ quark and $W$
boson. The $t'bW$ coupling
in general has the same form of our general $tbW$ coupling, 
and the expressions for single-$t'$ production cross sections are exactly
the same as single-top except for the heavy mass $m_{t'}$.
The size of the coefficients in the production cross sections 
decrease drastically with a greater
mass $m_{t'}$.  In Fig.~\ref{coefficient} we show their dependence
with respect to $m_{t'}$.  For instance, at $m_{t'}\,=500$ GeV
the $a_0$ coefficient decreases one order of magnitude with respect
to the value for $m_{t'}\,=178$ GeV.
Furthermore, in the t-channel single-$t'$ process, the 
$a_0$ coefficient, corresponding to longitudinal $W$ boson contribution,
dominates its production cross section.

\vspace{0.5cm}

\section{Models of EWSB}
For the second part of this paper, we would like to illustrate how
this approach can be used to make distinction among different
models of EWSB beyond the SM.   For simplicity, we assume
that no right-handed bottom quark couplings are present,
i.e. $f_1^R\simeq 0$, $f_2^L\simeq 0$.  Thus, we only need
two observables, like $f_0$ and $\sigma_t$, to make our analysis.

At this time it is convenient to notice that $f_0$ will not
depend on $\epsilon_L (\equiv f_1^L-1)$ if the other three couplings
are zero.  In our simplified scenario, if $f_0$ (and $f_-$) departs
from the SM prediction then $f_2^R$ cannot be zero.
In fact, the sign of $\Delta f_- \equiv f_- - f_-^{SM}$ is fixed by
the sign of $f_2^R$.

We would like to consider two models in particular:
\begin{itemize}
\item The Minimal Supersymmetric Standard Model (MSSM)
with $\tan \beta >1$ studied in Ref.~\cite{oakes1}, and
\item the Topcolor assisted Technicolor model (TC2)
considered in Ref.~\cite{qiao}.
\end{itemize}

Let us start with the case of the MSSM discussed in
Ref.~\cite{oakes1}.  Concerning the $W$-polarization in $t\to bW$
decay, Electroweak-Supersymmetry (SUSY) and QCD-SUSY corrections
are of order a few per-cent and tend to cancel each other.
The overall effect is to increase the left-handed decay mode
at the expense of reducing the longitudinal mode.  Thus, for
most of the SUSY parameter space the prediction is for a positive
$f_2^R$.  It is not true that $f_2^R$ must be positive for all of
the MSSM parameter space, but we can consider the positive sign of
this coupling as an indication of some scenarios of MSSM~\cite{oakes1}.

As for the second model, the TC2 scalars that couple strongly
with the top quark will modify the $tbW$ vertex in such a way as
to reduce $f_-$ in favor of $f_0$~\cite{qiao}.  This means that
in this case the sign of $f_2^R$ must be negative.

From the above discussion we can see that these two models have
a general tendency to predict opposite signs for the coupling
$f_2^R$.  The size and
sign of the other coefficient $\epsilon_L$ may depend on the
corresponding set of parameters of each model, let us assume
the following values as representative of each model:
\bea
{\rm MSSM:} \;\; \epsilon_L &=& 0.01\, ,
\;\;\;\;\; f_2^R = 0.005\, ,  \nonumber \\
{\rm TC2:} \;\; \epsilon_L &=& -0.01\, ,\;\; f_2^R = -0.005\, ,
\label{choices}
\eea
These numerical values were chosen such that the predictions for
the observables are consistent with
the results shown in Refs.~\cite{oakes1,qiao}.
(In the TC2 model, the size of the allowed $\epsilon_L$ 
and $ f_2^R$ could be much larger \cite{qiao}.)
Here, we ignore the $q^2$ dependence of the form factors.  This is
a reasonable approximation for the study of $t\to bW$.  Furthermore,
$\sigma_t$ comes predominantly from the small region of the
invariant mass of the $t\bar b$ pair, where the variation on $q^2$
can be ignored.

\begin{table}[ht]
\begin{tabular}{|c||c|c|}
\hline   & MSSM & TC2 
\tabularnewline \hline \hline
$\epsilon_L$ & 0.01 & -0.01   
\tabularnewline \hline 
$f_2^R$ & 0.005 & -0.005 
\tabularnewline \hline \hline 
${\Delta f_0}/{f^{SM}_0}$ &
$-0.5\%$ & $0.5\%$  
\tabularnewline \hline 
${\Delta f_-}/{f^{SM}_-}$ &
$1.2\%$ & $-1.2\%$   
\tabularnewline \hline 
(Tevatron) ${\Delta \sigma_t}/{\sigma^{SM}_t}$ & 
$2.1\%$ & $-2.0\%$  
\tabularnewline \hline 
(Tevatron) ${\Delta \sigma_s}/{\sigma^{SM}_s}$ & 
$3.2\%$ & $-3.1\%$  
\tabularnewline \hline 
(LHC) ${\Delta \sigma_t}/{\sigma^{SM}_t}$ & 
$2.2\%$ & $-2.1\%$  
\tabularnewline \hline 
(LHC) ${\Delta \sigma_s}/{\sigma^{SM}_s}$ & 
$3.4\%$ & $-3.3\%$  
\tabularnewline \hline 
${\Delta \Gamma_t}/{\Gamma^{SM}_t}$ & 
$3.5\%$ & $-3.4\%$ 
\tabularnewline \hline
\end{tabular}
\caption{Different model predictions for $f_0$, $f_-$, $\sigma_t$,
$\sigma_s$ and $\Gamma_t$.  Production of $\bar t$ is not included.
\label{models}}
\end{table}

In Table~\ref{models} we show the predictions of the two models
on the proposed observables.
Here, we do not include possible new production channels for the
s-channel single top events.  For example, it can be produced from
a $W'$ resonance whose contribution to $\sigma_s$ depends on the
other parameters of the model.  Nevertheless, the t-channel production
rate $\sigma_t$ is less sensitive to the other parameters because the
heavy resonance state contribution is suppressed by its large
mass.  Therefore, we shall concentrate on the measurements of $f_0$
and $\sigma_t$ in the following.

\begin{figure}
\includegraphics[width=8.2cm,height=6.4cm]{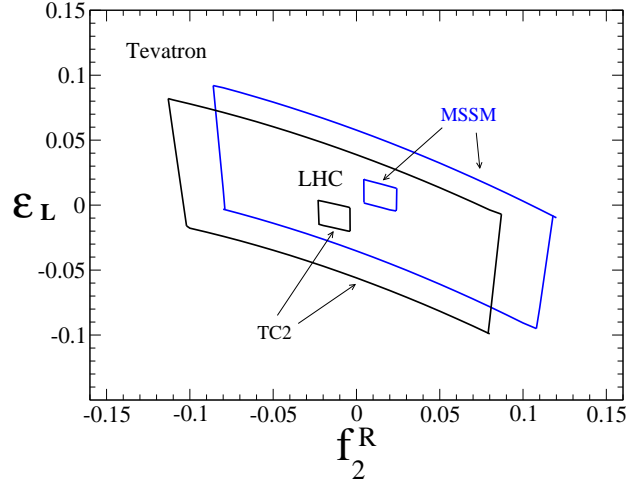}
\caption{Possible scenarios and the allowed $f_2^R$ vs $\epsilon_L$
region as given by measurements at the Tevatron and the LHC.}
\label{region}
\end{figure}

In Fig.~\ref{region} we show the sensitivity of the Tevatron and the
LHC to the determination of the couplings $\epsilon_L$ and $f_2^R$
for the above two model scenarios.  We assume that $f_0$
($\sigma_t$) can be measured to $10\%$ ($10\%$) accuracy at the
Tevatron, and to $1\%$ ($2\%$) accuracy at the LHC~\cite{review}.  
As for the LHC potential to measure single top production, the
CKM matrix element $V_{tb}$ could be measured down to less than one
percent error (statistical error only) at the ATLAS detector \cite{parsons}.
We conclude that the MSSM and TC2 could be distinguished
from each other at the LHC, but not at the Tevatron.

We want to emphasize that in general all four observables of
Eqs.~(\ref{eqf0})-(\ref{sigmas}) are needed to determine the
four couplings of the $tbW$ vertex and to make a complete
analysis that could test the different models of EWSB.

\noindent
{\bf Acknowledgments}~~~
We thank A. Belyaev and  H.-J. He for discussions.  The work of
F.L. has been supported in part by the Fulbright-Garcia Robles
grant and by Conacyt.  This work was supported in part by the NSF
grant PHY-0244919.

\end{document}